\documentclass[a4paper]{jpconf}
\usepackage{graphicx}
\begin{document}
\title{Nonequilibrium dynamics of polariton entanglement in a cluster of coupled traps}

\author{L.Quiroga$^1$ and C.Tejedor$^2$}

\address{$^1$Departamento de F\'{\i}sica, Universidad de Los Andes,
A.A.4976, Bogot\'a D.C., Colombia}
\address{$^2$Departamento de F\'{\i}sica Te\'orica de la Materia
Condensada, Universidad Aut\'onoma de Madrid, Cantoblanco, E-28049,
Madrid, Spain}

\ead{lquiroga@uniandes.edu.co}

\begin{abstract}
We study in detail the generation and relaxation of quantum
coherences (entanglement) in a system of coupled polariton traps. By
exploiting a Lie algebraic based super-operator technique we provide
an analytical exact solution for the Markovian dissipative dynamics
(Master equation) of such system which is valid for arbitrary
cluster size, polariton-polariton interaction strength, temperature
and initial state. Based on the exact solution of the Master
equation at $T=0$K, we discuss how dissipation affects the quantum
entanglement dynamics of coupled polariton systems.
\end{abstract}

\section{Introduction}
Coupled bosonic systems appear naturally in a great variety of
physical situations ranging from quantum optics, atomic/molecular
physics to condensed matter physics. In the last field
exciton/polariton systems are among the most notable examples.
Similarly to ultracold boson atoms confined in optical lattices,
trapped exciton/polariton systems are predicted to present different
collective phases such as Mott insulator or superfluid phases.
Experimental control of polariton trapping is rapidly improving.
Among the most promising candidates are stress induced trap systems
\cite{balili} and metal thin-film deposition on microcavities
\cite{kim}. A proposal for the generation of polarization entangled
photon-pairs from electric gate controlled polariton trapped systems
has been recently put forward \cite{na}. However, for full
exploitation of the capabilities of semiconductor microcavities
based entanglement, the polariton entangling power itself requires
complete characterization. In the present work, we focus on
polariton trapped clusters as open quantum systems evolving under
nonequilibrium or dissipative conditions. The time evolution of the
entanglement generated by polariton-polariton interactions jointly
with dissipative effects is considered.

In order to analyze the dissipative quantum dynamics of a trapped
polariton system we use a Born-Markov reduced dynamics approach. The
solution of the resulting Master (Lindblad) equation (ME) describes
the time evolution of the polariton system of interest. While
deviations from a strictly Markovian behavior may occur in real
microcavities \cite{prb08}, our analytical results may be used as a
first step in exploring more complex relaxation dynamics in
multitrap polariton systems. We restrict ourselves to the case where
the system is deep in the Mott insulator or fully localized regime
where polariton hopping is neglected but polariton-polariton
interactions remain important. To find the full density matrix of
the coupled polariton system, we use an exact Lie algebraic
superoperator method for solving the ME, generalizing previous
similar treatments \cite{ban,moya2} to arbitrary number of coupled
systems. By applying these results to a double trap, we focus our
attention to noise related properties of the entanglement creation
and preservation in such condensed matter highly nonlinear systems.
We quantify the amount of entanglement in terms of the logarithmic
negativity which has been proven to be valid as an entanglement
measure for any mixed state of arbitrary bipartite systems
\cite{vidal}. Furthermore, this entanglement measure is easily
computable having additionally an operational interpretation. We
found that in the Mott regime, the polariton-polariton interaction
generates naturally entanglement which is degraded by dissipation on
a shorter time scale as compared with the population decay, and with
an entanglement maximum being an increasing function of the initial
number of polaritons. The entanglement duration time shortens as the
polariton decay rates or the initial trap populations become large.

\section{Theoretical method}
Within the Born-Markov approach, the ME for the reduced system's
density operator $\hat{\rho}$ can be written as ($\hbar=1$)
\begin{eqnarray}
\frac{d\hat{\rho}(t)}{dt}&=&-i
[\hat{H},\hat{\rho}(t)]-\hat{\cal{L}}\hat{\rho}(t) \label{Eq:q1}
\end{eqnarray}
with the first term on the right hand side describing the coherent
or unitary evolution term, whereas the second term, or Lindblad
term, is associated to the noisy dynamics induced by the reservoirs.
Thus, we address the study of the relaxation dynamics of $M$ coupled
polariton traps described by the Hamiltonian
\begin{eqnarray}
\hat{H}=\sum_{n=1}^M \omega_n \hat{a}_n^{\dag}\hat{a}_n
+\sum_{n=1}^M\sum_{m=1}^M\lambda_{n,m}
\hat{a}_n^{\dag}\hat{a}_n\hat{a}_m^{\dag}\hat{a}_m\label{Eq:q2}
\end{eqnarray}
where $\hat{a}_n^{\dag}$ ($\hat{a}_n$) are boson creation
(annihilation) operators in trap $n$. The coupling matrix
$\bar{\lambda}=\{ \lambda_{n,m} \}$ is a real symmetric matrix
(hereafter diagonal terms will be denoted simply as
$\lambda_{n,n}=\lambda_n$). Lindblad terms, as taken to describe
reservoirs acting independently on each polariton trap, are given by
\begin{eqnarray} \hat{{\cal{L}}}\hat{\rho}(t)=\sum_{n=1}^M \left ( D_n\left [ \hat{a}_n^{\dag}\hat{a}_n\hat{\rho}(t)+
\hat{\rho}(t)\hat{a}_n^{\dag}\hat{a}_n-2\hat{a}_n\hat{\rho}(t)\hat{a}_n^{\dag}\right
] +P_n \left [
\hat{\rho}(t)\hat{a}_n\hat{a}_n^{\dag}+\hat{a}_n\hat{a}_n^{\dag}\hat{\rho}(t)-2\hat{a}_n^{\dag}\hat{\rho}(t)\hat{a}_n\right
] \right ) \label{Eq:q3}
\end{eqnarray}
where coefficients $D_n$ and $P_n$ are related to the dissipation
and pumping mechanisms, respectively, for the $n$-th trap. In this
equation, there are no cross terms of the type
$\hat{a}_n^{\dag}\hat{a}_m$ because we consider that decay and
pumping on one trap is independent from other traps.

In order to evaluate the time evolution of the polariton traps
density operator we proceed as follows. We define super-operators in
the following form
\begin{eqnarray} \hat{K}^{(n)}_- \hat{O} &=&\hat{a}_n^{\dag}\hat{O}\hat{a}_n\quad ,\quad \hat{K}^{(n)}_+ \hat{O}
=\hat{a}_n\hat{O}\hat{a}_n^{\dag}\quad ,\quad \hat{K}^{(n)}_0
\hat{O} =
-\frac{1}{2}(\hat{a}_n^{\dag}\hat{a}_n\hat{O}+\hat{O}\hat{a}_n^{\dag}\hat{a}_n+\hat{O})\label{Eq:q4}
\end{eqnarray}
and $\hat{{\cal N}}_{n}\hat{O}=[\hat{a}_n^{\dag}\hat{a}_n,\hat{O}]$,
where $\hat{O}$ denotes any polariton operator. Furthermore, given
the fact that $
[\hat{a}_n^{\dag}\hat{a}_n\hat{a}_m^{\dag}\hat{a}_m,\hat{\rho}(t)]=-\left
[ \hat{{\cal N}}_n\left ( \hat{K}_0^{(m)}+\frac{1}{2}\right )
+\hat{{\cal N}}_m\left ( \hat{K}_0^{(n)}+\frac{1}{2}\right )\right
]\hat{\rho}(t)$, the ME for the coupled polariton system,
Eq.(\ref{Eq:q1}), can be written as
\begin{eqnarray} \frac{d\hat{\rho}(t)}{dt}&=&-i\sum_{n=1}^M   \tilde{\omega}_n\hat{{\cal
N}}_n \hat{\rho}(t)+2\sum_{n=1}^M\left [
\hat{\Omega}_n\hat{K}^{(n)}_0+D_n\hat{K}^{(n)}_++P_n\hat{K}^{(n)}_-+\frac{D_n-P_n}{2}\right
] \hat{\rho}(t) \label{Eq:q7}
\end{eqnarray}
with $\tilde{\omega}_n=\omega_n-\lambda_{n}-\sum_{m\neq
n}^M\lambda_{n,m}$ and
$\hat{\Omega}_n=D_n+P_n+i\lambda_{n}\hat{{\cal N}}_n+i\sum_{m\neq
n}^M\lambda_{n,m}\hat{{\cal N}}_m$. As the super-operators
$\hat{{\cal N}_n}$ commute with any other super-operator, i.e.
$[\hat{{\cal N}}_m,\hat{{\cal N}}_n]=[\hat{{\cal
N}}_m,\hat{K}^{(n)}_{\pm}]=[\hat{{\cal N}}_m,\hat{K}^{(n)}_0]=0$,
the solution to Eq.(\ref{Eq:q7}) has the form
\begin{eqnarray}
\hat{\rho}(t)=\prod_{n=1}^Me^{-i \tilde{\omega}_n\hat{{\cal N}}_n t}
e^{2\left [
\hat{\Omega}_n\hat{K}^{(n)}_0+D_n\hat{K}^{(n)}_++P_n\hat{K}^{(n)}_-+\frac{D_n-P_n}{2}\right
]t}\hat{\rho}(0) \label{Eq:q8}
\end{eqnarray}
where $\hat{\rho}(0)$ denotes the initial state. It is
straightforward to demonstrate that the superoperators defined by
Eq.(\ref{Eq:q4}) close in the su(1,1) Lie algebra
\begin{eqnarray}
[\hat{K}^{(m)}_+,\hat{K}^{(n)}_-]=-2\hat{K}^{(n)}_0\delta_{m,n}\quad
,\quad [\hat{K}^{(m)}_0,\hat{K}^{(n)}_{\pm}]=\pm
\hat{K}^{(n)}_{\pm}\delta_{m,n} \label{Eq:q5}
\end{eqnarray}
By using the su(1,1) Lie algebra structure of the superoperators as
given in Eq.(\ref{Eq:q5}), the ME solution expressed by
Eq.(\ref{Eq:q8}), can finally be written in the full disentangled
form
\begin{eqnarray}
\hat{\rho}(t)=\prod_{n=1}^Me^{-i \tilde{\omega}_n\hat{{\cal N}}_n t}
e^{(D_n-P_n) t}
e^{\hat{c}_-^{(n)}(t)\hat{K}^{(n)}_-}e^{\hat{c}^{(n)}_0(t)\hat{K}^{(n)}_0}e^{\hat{c}_+^{(n)}(t)\hat{K}^{(n)}_+}\hat{\rho}(0)
\label{Eq:q9}
\end{eqnarray}
where the temporal coefficients $\hat{c}^{(n)}_{\pm}(t)$ and
$\hat{c}^{(n)}_0(t)$ are functions of $\hat{\Omega}_n$ (details are
presented elsewhere \cite{uam1}). In this way any expectation value
of polariton observables could be analytically evaluated.

\section{Entanglement nonequilibrium dynamics}

Given the general solution of the ME in Eq.(\ref{Eq:q9}) an
analytical expression can be obtained for the cluster density
operator \cite{uam1}. However, in the present work we focus on the
zero temperature case for a two trap polariton cluster ($M=2$). The
same algebraic results could also be used to model a single trap
with two polariton species, as it could be the case for a system
formed by polaritons of different spin polarizations confined in the
same spatial region. We consider that the system is initially
prepared, by a resonant short laser pulse, in a purely separate
coherent state, i.e..
$\hat{\rho}(0)=|\alpha_1,\alpha_2><\alpha_1,\alpha_2|$, where
$|\alpha_j>$ denotes a coherent state in the $j$-th trap, $j=1,2$.
The initial mean number of polaritons in trap $j$ is given by
$|\alpha_j|^{2}$. After this initialization step, interacting
polaritons evolve in time in presence of dissipation while, in the
vanishing temperature limit, incoherent pumping effects are not
present ($P_n=0$). Moreover, we take identical dissipation decay
rates $D_n=\Gamma$. We take as the unit of energy the inter-trap
interaction strength $\lambda_c=1$ and consequently time is measured
in units of $\lambda_c^{-1}$.

Now the question is: how to identify entanglement between two
bosonic systems?. By contrast with two qubit systems for which
entanglement can be unambiguously identified by the concurrence
\cite{wootters}, for two continuous variables as it is the case
here, sufficient and necessary conditions for detecting entanglement
have only been established for the special case of Gaussian states
\cite{duan,simon}. However, in general the polariton trap state is a
mixed non-Gaussian state.

We quantify the polariton entanglement by computing the logarithmic
negativity $LN(\rho)=log_2||\rho^{PT}||$, where $||\rho^{PT}||$ is
the trace norm of the partial transpose density operator
\cite{vidal}. This latter quantity can be easily evaluated as
$||\rho^{PT}||=2N(\rho)+1$ where $N(\rho)$, or negativity, is the
absolute value of the sum of the negative eigenvalues of
$\rho^{PT}$. Based on the exact analytical solution for the ME given
in Eq.(\ref{Eq:q9}), the required transpose operation on the
system's density operator is trivial for any time from which the
negative eigenvalues of $\rho^{PT}$ are numerically evaluated. For
any separable or unentangled state $LN(\rho)=N(\rho)=0$. Obviously,
the entanglement fully vanishes for $\lambda_c=0$. To focus the
discussion and for demonstrational purposes we consider here the
case of identical traps, i.e. $\omega_1=\omega_2$ and
$\lambda_1=\lambda_2$, which are also identically prepared
$\alpha_1=\alpha_2$.

\begin{figure}[h]
\begin{minipage}{14pc}
\includegraphics[width=6cm,angle=-90]
{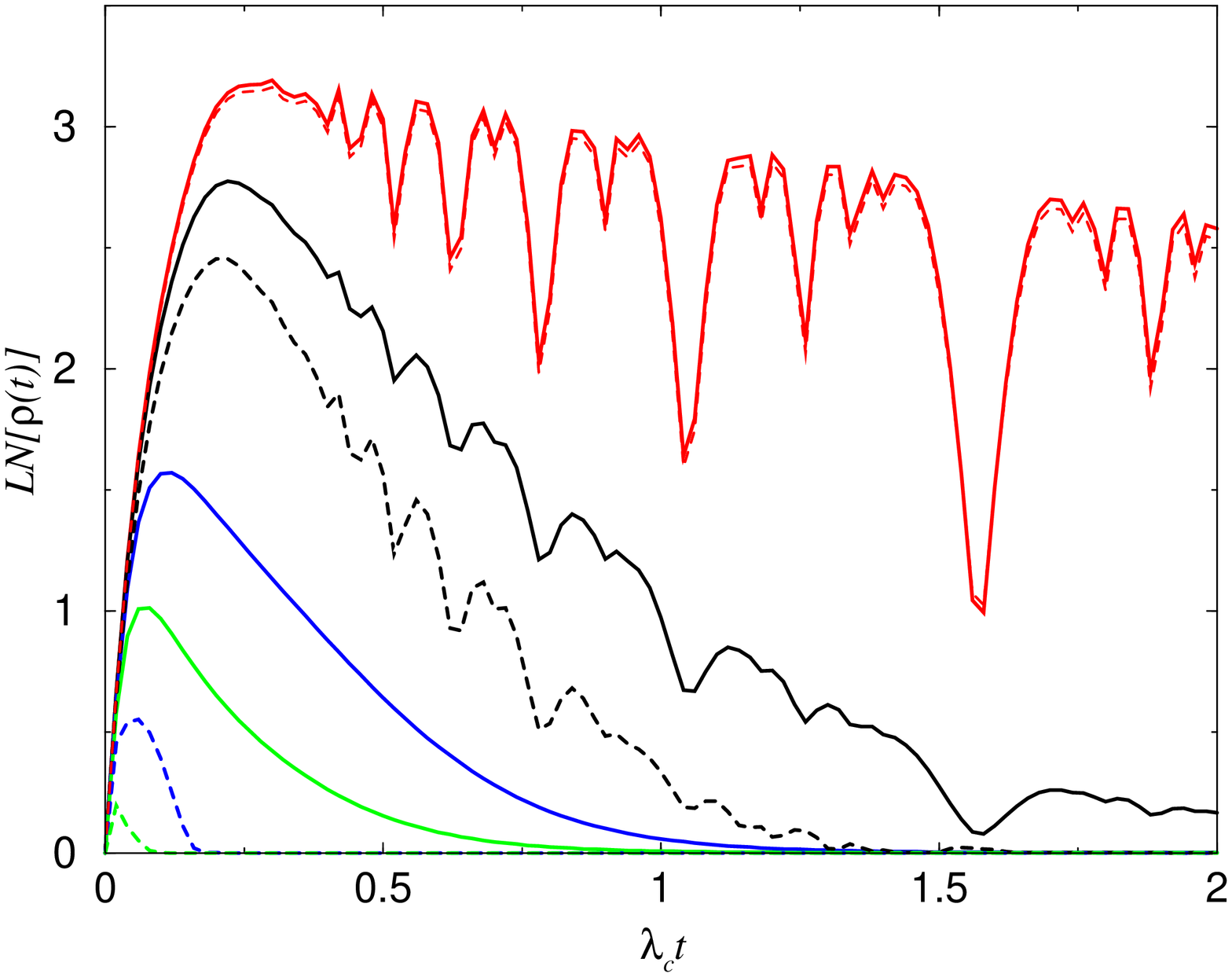} \caption{\label{label}Logarithmic negativity as a
function of time for different polariton self-interaction strengths
and dissipation rates. See text for parameters.}
\end{minipage}
\hspace{6pc}
\begin{minipage}{14pc}
\includegraphics[width=6cm,angle=-90]{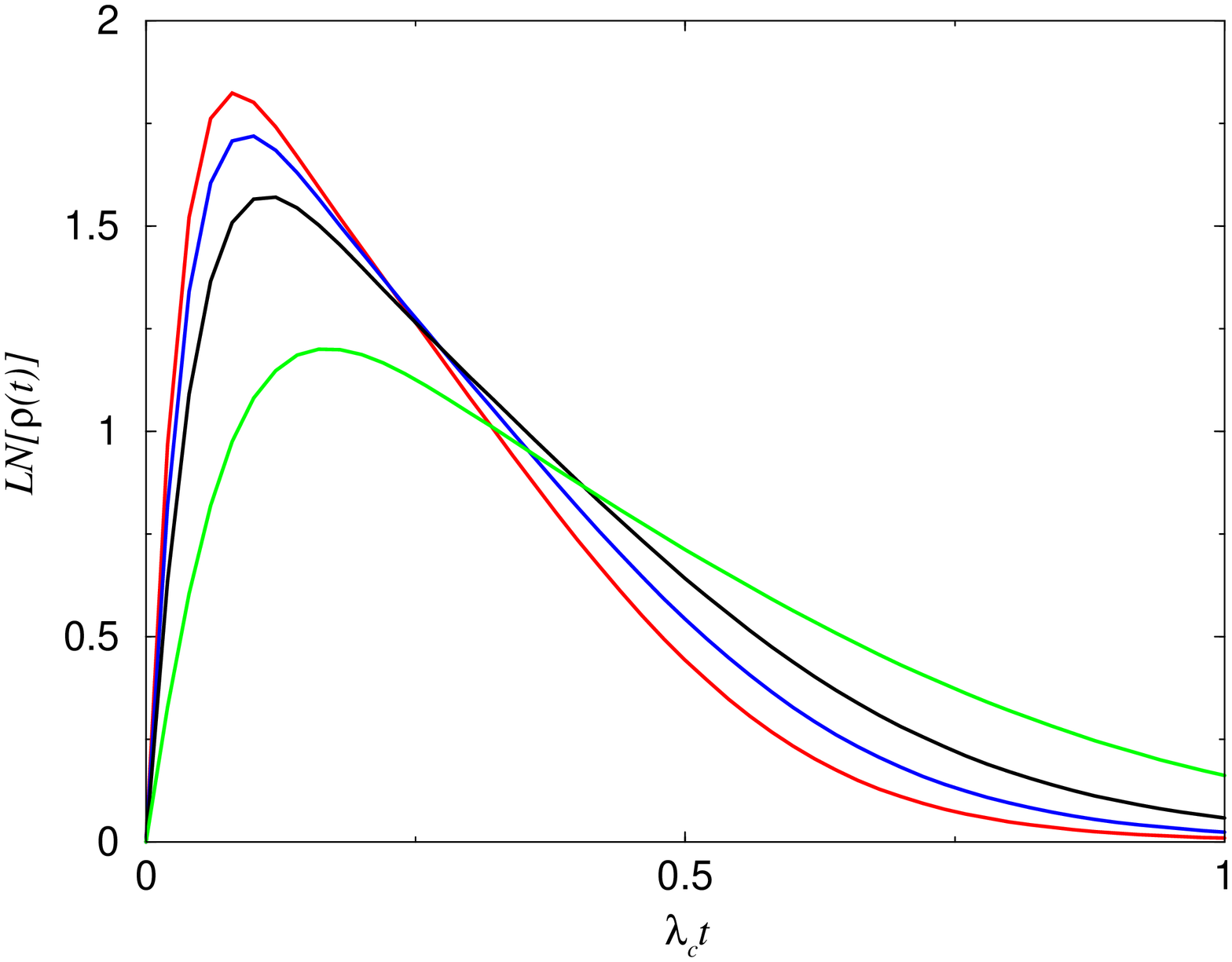}
\caption{\label{label}Logarithmic negativity as a function of time
for different polariton populations. See text for parameters.}
\end{minipage}
\end{figure}

In Figure (1) we plot the time evolution of the polariton
entanglement, $LN[\rho(t)]$, for a two trap cluster initially
prepared with $|\alpha_j|^{2}=6$ (the amount of entanglement does
not depend on the frequency $\omega_j$ of each trap). Four different
values of dissipation rates $\Gamma=0.01, 0.1, 1$ and $3$ are
considered and correspond in Fig.(1) to colors red, blue, black and
green, respectively. Additionally, for each decay rate, results for
two extreme values of the ratio between the intra-trap and
inter-trap interactions are displayed: $\lambda_j/\lambda_c=1$ solid
lines and $\lambda_j/\lambda_c=10$ dashed lines (colors are still as
described above). The first observation is that entanglement is
immediately generated by the polariton-polariton interaction,
$\lambda_c$, and it is preserved for a significant fraction of the
time that polariton population in the traps is maintained. The
cluster system starts its evolution in a separable state rising its
entanglement degree in an almost linear way with a slope depending
directly on the initial polariton population, a clear indication of
how important is the effective non-linear polariton interactions for
generating entanglement. At long times the system ends again in a
separable state, as it should be because the traps become empty and
thus the final state is nothing but the double vacuum state. At very
low decay rates entanglement oscillations are visible due to the
quasi-unitary evolution of inter-trap correlations governed
exclusively by $\lambda_c$ \cite{uam1}. Note that in this limit of
weak dissipation the entanglement dynamics does not depend on the
intra-trap polariton interactions $\lambda_j$, as it is evident from
the almost identical solid and dashed red curves in Fig.(1). As the
dissipation becomes important populations and the entanglement decay
faster. The polariton population follows a very simple exponential
decay form given by
$<\hat{a}^{\dag}_j\hat{a}_j>(t)=|\alpha_j|^{2}e^{-2\Gamma t}$ while
for the entanglement a more pronounced dependence on the polariton
self-interactions are evident. Thus, dissipation and polariton
self-interactions jointly degrade significantly both the maximum
attainable entanglement and the time interval where entanglement is
visible.

Figure (2) shows the time evolution of $LN(\rho)$ for different
polariton initial populations $|\alpha_j|^{2}=10$ (red curve),
$|\alpha_j|^{2}=8$ (blue curve), $|\alpha_j|^{2}=6$ (black curve)
and $|\alpha_j|^{2}=3$ (green curve). Other parameters are:
$\lambda_1=\lambda_2=-1$ and $\Gamma=1$. It is evident that the
rising time of the entanglement is shorter as the polariton
population increases, emphasizing previous comments on the
importance of inter-trap polariton-polariton interactions for an
efficient and rapid generation of entanglement. However, on the
other hand, the temporal width of the entanglement curve decreases
as a function of the polariton population, indicating the high level
of non-linear effects occurring in the present system.

Some numerical estimates are in order. According to Ref.\cite{kim}
typical engineered trap radius range from $5$ to $50$ $\mu m$.
Pumping pulse lasers of low intensity on the order of $1$ $mW$ give
rise to a polariton density of $10^8-10^{10}$$cm^{-2}$. Thus, for
experimentally achievable trap sizes, typically $10-100$ polaritons
could be found in a single small trap. Our unit of time is fixed by
the inter-trap (or inter-species) interaction strength $\lambda_c$.
According to Ref.\cite{na}, $\lambda_c\sim 10-100$ $\mu eV$, which
for $\lambda_c t\sim 1$ yields to $t\sim 40$ $psec$. Hence, the
theoretical results we discuss above are in the lower range of
experimentally accessible polariton densities and time resolution,
requiring very clean microcavities samples and high quality factors
$Q$, for entanglement detection to be observable. However, our
analytical results should be indeed valuable as a starting point for
more numerically involved calculations in realistic situations.

\section{Conclusions}

By solving exactly the Master equation for an interacting trapped
polariton system we have found that polariton entanglement can be
immediately produced from an initial separable multi-coherent state.
The entanglement starts rising linearly with a slope depending on
the initial trap populations. The entanglement maximum is an
increasing function of the initial number of polaritons. However,
the entanglement duration time shortens as the initial trap
population becomes large. In any case polariton entanglement
observation would require very good microcavities with weak
dissipation rates.

%\subsection{Acknowledgments}
LQ acknowledges MEC(Spain) for a sabbatical grant and Universidad
Aut\'onoma de Madrid-Spain for hospitality during the development of
the present work. This work has been supported in part by the
Spanish MEC under contract QOIT Consolider-CSD2006-0019 and by CAM
under contract S-0505/ESP-0200.

\end{document}